\documentclass[prl,twocolumn,superscriptaddress,fixfloat]{revtex4}
\usepackage{amsmath}
\usepackage{amssymb}
\usepackage{graphicx}
\usepackage{dcolumn}

\def \bise{Bi$_2$Se$_3$}

\begin{document}

\title{Bulk Fermi surface coexistence with Dirac surface state in Bi$_2$Se$_3$: a comparison of photoemission and Shubnikov-de Haas measurements}

\author{James G. Analytis}
\affiliation{Stanford Institute for Materials and Energy Sciences, SLAC National Accelerator Laboratory, 2575 Sand Hill Road, Menlo Park, CA 94025, USA} \affiliation{Geballe Laboratory for Advanced Materials and Department of Applied Physics, Stanford University, USA}

\author{Jiun-Haw Chu}
\affiliation{Stanford Institute for Materials and Energy Sciences, SLAC National Accelerator Laboratory, 2575 Sand Hill Road, Menlo Park, CA 94025, USA} \affiliation{Geballe Laboratory for Advanced Materials and Department of Applied Physics, Stanford University, USA}

\author{Yulin Chen}
\affiliation{Stanford Institute for Materials and Energy Sciences, SLAC National Accelerator Laboratory, 2575 Sand Hill Road, Menlo Park, CA 94025, USA} \affiliation{Geballe Laboratory for Advanced Materials and Department of Applied Physics, Stanford University, USA}

\author{Felipe Corredor}
\affiliation{Stanford Institute for Materials and Energy Sciences, SLAC National Accelerator Laboratory, 2575 Sand Hill Road, Menlo Park, CA 94025, USA} \affiliation{Geballe Laboratory for Advanced Materials and Department of Applied Physics, Stanford University, USA}

\author{Ross D. McDonald}
\affiliation{Los Alamos National Laboratory, Los Alamos, NM 87545, USA}

\author{Z. X. Shen}
\affiliation{Stanford Institute for Materials and Energy Sciences, SLAC National Accelerator Laboratory, 2575 Sand Hill Road, Menlo Park, CA 94025, USA} \affiliation{Geballe Laboratory for Advanced Materials and Department of Applied Physics, Stanford University, USA}

\author{Ian R. Fisher}
\affiliation{Stanford Institute for Materials and Energy Sciences, SLAC National Accelerator Laboratory, 2575 Sand Hill Road, Menlo Park, CA 94025, USA} \affiliation{Geballe Laboratory for Advanced Materials and Department of Applied Physics, Stanford University, USA}

%\author{I. R. Fisher}
%\affiliation{Geballe Laboratory for Advanced Materials and Department of Applied Physics, Stanford University}

\begin{abstract}
Shubnikov de Haas (SdH) oscillations and Angle Resolved PhotoEmission
Spectroscopy (ARPES) are used to probe the Fermi surface of single
crystals of \bise. We find that SdH and ARPES probes quantitatively
agree on measurements of the effective mass and bulk band
dispersion. In high carrier density samples, the two probes also agree
in the exact position of the Fermi level $E_F$, but for lower carrier
density samples discrepancies emerge in the position of $E_F$. In
particular, SdH reveals a bulk three-dimensional Fermi surface for
samples with carrier densities as low as 10$^{17}$cm$^{-3}$. We
suggest a simple mechanism to explain these differences and discuss
consequences for existing and future transport studies of topological
insulators.
\end{abstract}


\maketitle

 Recently, a new state of matter, known as a topological insulator,
 has been predicted to exist in a number of materials:
 Bi$_{1-x}$Sb$_x$, Bi$_2$Se$_3$, Bi$_2$Te$_3$ and
 Sb$_2$Te$_3$\cite{teo_surface_2008,zhang_topological_2009}. This
 state of matter is characterized by a full band gap in the bulk of
 the material, but with a gapless, dissipationless surface state. The
 surface state is comprised of counter-propagating spin states, which
 create a dispersion of a single, massless Dirac cone that is
 protected by time-reversal symmetry. The experimental realization of
 this state could mean significant advances in spintronic devices,
 quantum computation and much more besides.

\begin{figure}
\includegraphics[width = 9.2cm ]{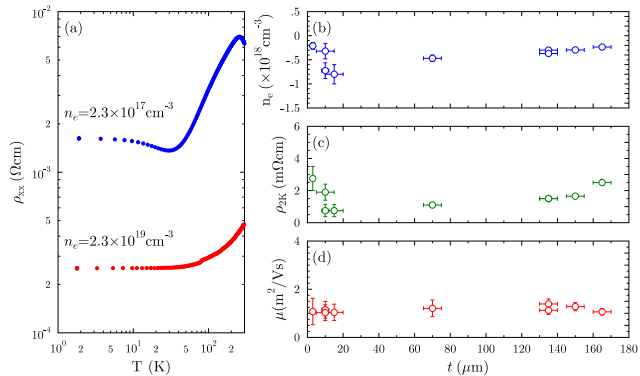}
\caption{(a)Temperature dependence of two typical samples of
  Bi$_2$Se$_3$ with carrier densities differing by two orders of
  magnitude. (b) Shows the carrier density $n_e$, (c) the resistivity
  $\rho_0$ at T=2K and (d) the mobility for samples of different
  thicknesses. Each sample was a cleave from a parent sample, so that
  the surface area of each sample was kept constant.}
\label{res_temp} 
\end{figure}

As a result there has been great excitement in the last year after the
discoveries of various ARPES experiments
\cite{hsieh_observation_2009,chen_experimental_2009,hsieh_tunable_2009}
and more recently from scanning-tunneling measurements
\cite{alpichshev_stm_2009,roushan_topological_2009,gomes_quantum_2009}
that such a state appears to exist in nature. Amidst this flurry of
recent results, it is easy to forget that these same materials have
been the subject of careful and thorough research for much of the 20th
century. However, common to all the unambiguous measurements of the
Dirac cone is the use of surface-sensitive probes. Only recently have
transport measurements emerged specifically investigating the surface
state (Refs \cite{taskin_quantum_2009,peng_aharonov-bohm_2009,
  checkelsky_giant_2009}), all of which note the dominance of the bulk
conductivity.  It is thus of great interest to perform a coordinated
study of these materials using both bulk transport experiments and
surface sensitive ARPES experiments. Here we report results of these
investigations. The transport experiments reveal quantum oscillations
that indicate a bulk band structure and Fermi surface volume that
monotonically change with doping.  For carrier densities in the range
$\sim$10$^{19}$cm$^{-3}$, the transport extracted band structure is in
quantitative agreement with the bulk band structure determined by
ARPES which also observes the Dirac dispersion of the surface
state. The quantitative agreement between ARPES and SdH provides
additional support for the existence of novel band structure in these
materials. For lower carrier density samples down to
$10^{17}$cm$^{-3}$ we observe SdH oscillations which unambiguously pin
the Fermi level in the bulk conduction band, with a high level of
consistency across all samples measured from the same batch. While
ARPES places $E_F$ near the SdH level for some samples, there are
others from the same batch whose $E_F$ is found to reside into the
bulk gap. We discuss possible explanations for these discrepancies and
the implications for transport studies of surface Dirac Fermions in
samples near a metal-insulator transition.

The material Bi$_2$Se$_3$ can be grown without the introduction of
foreign dopants as either $n$ or $p$
type\cite{kohler_conduction_1973,kohler_galvanomagnetic_1975} though
is more commonly found as the former because the dominant defects tend
to be Se vacancies. Quantum oscillatory phenomena, which provides
evidence of bulk metallic behavior has been reported by Kohler {\it et
  al.}  \cite{kohler_conduction_1973} on low carrier density samples
and more recently by Kulbachinskii
\cite{kulbachinskii_conduction-band_1999} on high carrier density
samples. Below a carrier density of 7$\times 10^{18}$cm$^{-3}$, the
band structure is well approximated by a single parabolic band, making
the interpretation of transport measurements
transparent\cite{kulbachinskii_conduction-band_1999}. Two $n$-type
samples with carrier densities differing by two orders of magnitude
are shown in Figure \ref{res_temp}. For the low carrier density
samples an upturn in the resistivity is seen, which levels off at
sufficiently low temperature. This behavior has been attributed to the
presence of an impurity band whose thermally activated conductivity is
comparable to the band conductivity until carriers freeze out at
around 30K
\cite{kohler_galvanomagnetic_1975,kulbachinskii_conduction-band_1999}. This
behavior is not apparent in the higher carrier density materials,
where the band conductivity always dominates. Even though we have
reduced the carrier density by 2 orders of magnitude, the resistivity
increases by one, suggesting that the mobility has increased in the
low carrier density samples, consistent with previous measurements
\cite{kulbachinskii_conduction-band_1999}. The low carrier density
samples are around an order of magnitude smaller than those of
reported topological insulators including Sn doped
Bi$_2$Te$_3$\,($n_e\sim 1.7\times10^{18}$cm$^{-3}$)
\cite{chen_experimental_2009} or Ca doped Bi$_2$Se$_3$($n_e\sim
5\times10^{18}$cm$^{-3}$)\cite{hsieh_tunable_2009} and as a result may
be better candidates in which to observe the transport properties
dominated by the topological surface state. 

\begin{figure}
\includegraphics[width = 8.2cm ]{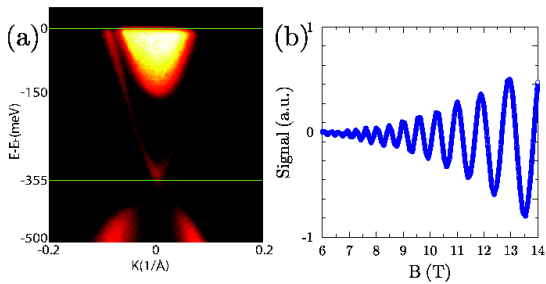}
\caption{(a) ARPES band dispersion on samples of Bi$_2$Se$_3$ with carrier
  density 2.3$\times 10^{19}$cm$^{-3}$ (batch S4). (b) Due to the
  quantization of the energy spectrum into Landau levels (LLs),
  oscillations appear in the magnetoresistance known as SdH
  oscillations. The SdH oscillations here are for a sample taken from
  the same batch as in (a) at $\theta=0$, corresponding to a
  oscillatory frequency of F$=155T$, consistent with $E_F\sim
  160$meV. ARPES and SdH are in good agreement for these high carrier
  density samples.}
\label{sdh_arpes_high} 
\end{figure}

Single crystals of Bi$_2$Se$_3$ have been grown by slow cooling a
binary melt. Elemental Bi and Se were mixed in alumina crucibles in a
molar ratio of 35:65 for batch S1 ($n_e=5\times10^{17}$), 34:66 for
batch S2 ($n_e=3\times10^{17}$), 34:66 for batch S3
($n_e=2.3\times10^{17}$), and 40:60 for batch S4
($n_e=2.3\times10^{19}$). The mixtures were sealed in quartz ampules
and raised to 750 $^\circ$C and cooled slowly to 550 $^\circ$C, then
annealed for an extended period. Crystals can be cleaved very easily
perpendicular to the (0 0 1) axis. Measurements of the resistivity and
Hall effect were measured in a 14T PPMS using a standard 4-probe
contact configuration and Hall measurements were performed using a
6-probe configuration. For the latter, only data which was linear in
the low field limit was used to avoid mixing with longitudinal
components. In addition to this precaution, signal from positive and
negative field sweeps was subtracted to extract the odd (Hall)
components of the signal, after which the carrier density is extracted
in the usual way. ARPES measurements were performed at beam line
10.0.1 of the Advanced Light Source (ALS) at Lawrence Berkeley
National Laboratory. Measurement pressure was kept
$<$3$\times$10$^{11}$ Torr, and data were recorded by Scienta R4000
analyzers at 15K sample temperature. The total convolved energy and
angle resolutions were 16meV and 0.2$^\circ$
(i.e. $<$0.007($\AA^{-1}$) or $<$0.012(1$\AA^{-1}$) for photoelectrons
generated by 48eV photons), at which energy the cross-section for both surface state
and bulk bands is strong.

In Figure \ref{sdh_arpes_high} we show complimentary ARPES and SdH
data on samples from the same batch, with carrier density determined
by the Hall effect of $n_e=2.3\times 10^{19}$cm$^{-3}$. The SdH
reveals an anisotropic pocket of frequency 155 T, corresponding to a
filling of around 160meV (the band structure is not parabolic at this
filling and so we assume similar band structure parameters as Kohler
{\it et al.}\cite{kohler_conduction_1973} characterizing similar
carrier density samples of \bise). ARPES results on samples from the
same batch, show the Fermi level 150meV above the bottom of the
conduction band in good quantitative agreement. Similarly, the
effective mass (see below) extracted from SdH is in good quantitative
agreement with that measured by ARPES.

\begin{figure}
\includegraphics[width = 9.2cm ]{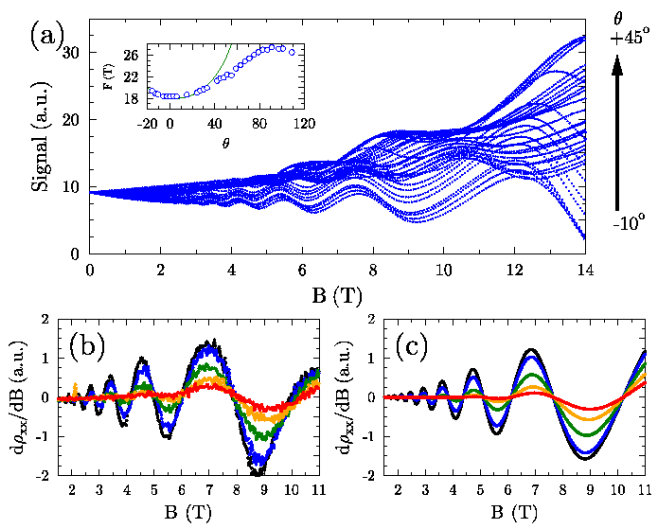}
\caption{ (a) Magnetotransport for samples from S1. As the angle is
  swept the frequency of the oscillation varies according to the
  topology of the Fermi surface. For a two-dimensional pocket the
  expected dependence is 1/cos$\theta$ (shown in green in the
  inset). The observed angle dependence is clear evidence for a closed
  ellipsoidal Fermi surface pocket, similar to that observed by Kohler
  {\it et al.}\cite{kohler_conduction_1973}. Similar SdH data was
  gathered on batch S2 and S3 on a number of samples. Samples from
  batch S3 showing the temperature dependence of the derivative of the SdH
  signal in (b) and a fit to the data shown in (c) from which the effective
  mass, Dingle temperature and oscillatory frequency can be
  extracted.}
\label{TI_SdH} 
\end{figure}

In Figure \ref{TI_SdH} we illustrate angle dependent SdH data (a)
taken at 1.8K, on a sample from batch S1 with a lower carrier density
of 10$^{17}$cm$^{-3}$. The SdH signal reveals a pocket that is
approximately an ellipsoid elongated about the c3 axis, consistent
with measurements by Kohler et al. from the
1970's\cite{kohler_conduction_1973} on samples with similar carrier
densities. For a two-dimensional pocket expected from the surface
state, quantum oscillations should vary as 1/cos$\theta$, where
$\theta$ is the angle between the c3 axis and the field direction, so
the present observations must originate from a 3D Fermi surface
existing in the bulk. It has been shown by Kohler {\it et al.} and
more recently by Kulbachinskii {\it et al.}  that the conduction band
structure for these low carrier densities is approximately
parabolic\cite{kohler_conduction_1973,kulbachinskii_conduction-band_1999},
and so the band filling can be estimated by
$E_F=\frac{\hbar^2A_k}{2\pi m^*}$,where $A_k$ is the area of the Fermi
surface in Fourier space. We estimate the Fermi energy to be 18meV
above the bottom of the conduction band.

In Figure \ref{TI_SdH} (b) we show the derivative of the longitudinal
magnetoresistance of a sample from batch S3 and a fit of the entire
data set using the usual Lifshitz-Kosevich formalism, to extract the
effective mass and Dingle temperature T$_D$, with fit shown in
(c). Fitting the entire data set, which is often more accurate than
tracing the amplitude of the Fourier transform, our fit yields
$m^*=0.15m_e$ and T$_D=3.5K$, for this frequency (F=14T). Similar data
for samples from batch S4 give $m^*=0.125m_e$, T$_D=4K$ and
F=155T. The mean free path is calculated using the orbitally averaged
velocity and scattering time extracted from T$_D$ yields
$l_{S3}\sim60$nm and $l_{S4}\sim220$nm. This data is wholly consistent
with the very complete SdH studies of Kohler {\it et
  al.}\cite{kohler_conduction_1973,kohler_galvanomagnetic_1975} and
more recently by Kulbachinskii {\it et al.}
\cite{kulbachinskii_conduction-band_1999}. In addition, the data was
reproduced with high consistency on a number of samples from the same
batch, and even on samples from different batches with similar growth
parameters.

\begin{figure}
\includegraphics[width = 9.2cm ]{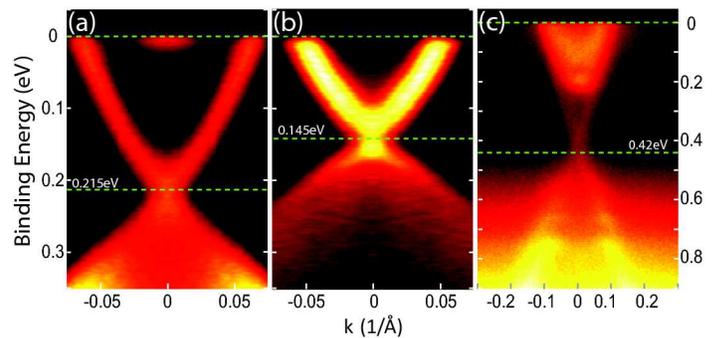}
\caption{ARPES data on samples of Bi$_2$Se$_3$ from batch S1. The horizontal
  lines show the crossing of the Fermi level ($E-E_F$ = 0) and the
  Dirac crossing. (a) Band structure measured by ARPES results on
  samples from batch S1 showed the Fermi level near the SdH level of
  $\sim 15meV$ from the bottom of the conduction band. (b) Measurement
  on another sample from batch S1 showed the Fermi level in gap. Some
  other samples from S2 and S3 also showed the Fermi level in the bulk
  gap. The variation might be due to the lower carrier density of
  these samples, and the surface band structure is more susceptible to
  small amounts of surface contamination. c) Band structure of a
  sample also from batch S1 which was cleaved in atmosphere and
  exposed for 10s, showing significant n-type doping with large bulk
  conduction band pocket.}
\label{TI_arpes} 
\end{figure}

ARPES data on samples from the same batches as those shown in Figure
\ref{TI_SdH}, determining the effective mass as $m^*-0.13$ in very
good quantitative agreement with SdH. However, the exact placement of
the Fermi level in the band structure reveals some disagreement. In
Figure \ref{TI_arpes} we illustrate photoemission data for two
separate samples from batch S1. In (a) the Fermi level is near the
bottom of the conduction band in agreement with SdH, while in (b) it
is in the gap (about 60meV below the conduction band), crossing the
Dirac cone with apparently no bulk contribution. A number of samples
from similar batches, such as batch S2 and S3, also have shown similar
variation in $E_F$ crossing the gap on some samples. While $E_F$
determined by photoemission appears to show some variation, it is
important to note that the measured $E_F$ from ARPES is either near or
below the SdH $E_F$.

Despite the good agreement of SdH and ARPES on high carrier density
samples, and the agreement of the effective mass and other band
parameters on the low carrier density samples, the discrepancy in the
position of the Fermi level requires explanation. Such differences can
occur for a number of reasons, for example due to sample variation
within a batch, or perhaps due to variation in the exposure of cleaved
surfaces before a photoemission measurement. However, it should be
noted that the SdH frequency does not appear to vary significantly
within a batch for up to 20 samples measured in the present study and
so the former seems an unlikely scenario. Another reason for the
discrepancy may be that atmospheric exposure of transport samples has
contaminated them with an $n$-type dopant causing them to appear bulk
$n$-type. Figure \ref{TI_arpes} (c) illustrates photoemission data for
a sample cleaved in air. The Dirac cone of the surface state remains
robust and the bulk conduction band appears partially occupied. Such
doping may lead to a 3D Fermi surface pocket appearing in SdH
oscillations if the contamination is deep enough and allows for
sufficiently long mean free paths.

To investigate this possibility further we measure the thickness
dependence of the transport by systematically thinning a single
sample.  Cleaving was achieved with tape, keeping the surface area of
the resulting samples relatively constant and allowing direct
comparison of data sets of each cleave. Though the samples are
vulnerable to deformations between cleaves, only data from mirror-like
flat samples is presented. In most cases these samples still exhibited
quantum oscillatory phenomena, confirming the high quality of the
cleaved samples. Figure \ref{res_temp} (b-d) shows a summary of the
low-temperature carrier density, resistivity and mobility. Within our
error bars, each quantity seems to vary weakly down to 3$\mu$m in
thickness. Although the residual resistivity and carrier density
varies slightly (possibly from disorder related to slight sample
deformation, despite the precautions mentioned above), the mobility
remains almost constant as a function of thickness at $\mu\sim
1$m$^2/$Vs. In summary, the transport is insensitive to the thickness,
suggesting that the SdH oscillations are not a consequence of
atmospheric contamination, but originate from the intrinsic band
conductivity of the bulk.

\begin{figure}
\includegraphics[width = 9.2cm ]{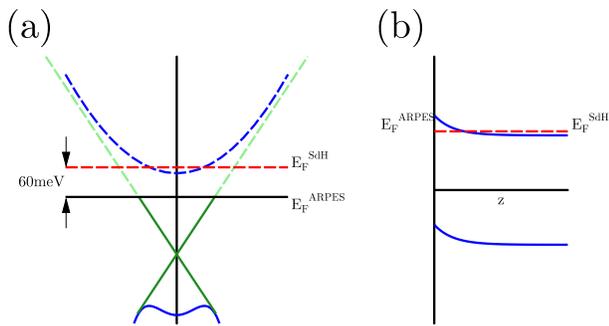}
\caption{(a) A schematic representation of the band structure seen by
  ARPES (solid) red horizontal line denoting the Fermi level as seen
  by SdH. (b) We infer band bending of about 60 meV at the surface
  from a comparison of ARPES and quantum oscillations.}
\label{band_bend} 
\end{figure}

A final scenario for the discrepancy is that the band structure is
distorted near the surface due to space-charge accumulation. This is
known to occur in many semiconductors, such as InSb or
CdTe\cite{king_surface_2008,swank_surface_1967}, whereby the bulk band
structure bends as the surface is approached. Typically, such bending
occurs over a surface depletion layer $z_d$, which can be calculated
by solving the Poisson equation to yield $z_d^2=\kappa\epsilon_0\Delta
V/en_e $\cite{mnch_semiconductor_2001}, where $\kappa$ is the DC
dielectric permittivity (estimated from these samples as $\sim 113$
\cite{landolt_bismuth_1998})and $\Delta V$ is the difference in energy
between the surface and bulk state. We estimate $z_{d}^{S1}\sim 40{\rm
  nm}$ for the low carrier density samples and $z_d^{S4}\sim 2{\rm
  nm}$ for the high carrier density samples. A schematic
representation of the band bending is shown in Figure
\ref{band_bend}. The present argument suggests that discrepancies
between ARPES and SdH can be explained, even expected for low carrier
density samples. In addition, due to the small value of $n$, these
samples are likely more susceptible to a small amounts of surface
contamination, especially if the uncontaminated surface $E_F$ is in
the gap (as illustrated in by ARPES on atmosphere exposed
samples). This may help explain why there is some variability in the
Fermi level of ARPES data but not in the SdH data.

Much theoretical work has emerged on the dramatic consequences of the
surface state on transport properties\cite{lee_surface_2009}. Yet over
several decades of experimental study, such properties have not been
observed. Recently, Aharanov-Bohm and universal conductance
fluctuations have been observed which may be due to the surface state
\cite{peng_aharonov-bohm_2009,checkelsky_giant_2009}, but even in
these cases the conductance appears bulk at the temperatures
considered. Conventionally, such intrinsically doped materials can
become `insulators' by either a Mott-like or an Anderson
transition. The first can occur when the Bohr radius
$a_B=\kappa\hbar^2/m^*e^2$ falls below the Thomas-Fermi screening
length $\lambda_{TF}$, so that wavefunctions cannot overlap. This can
be estimated using $\lambda_{TF}^2=\kappa\epsilon_0/(2\pi e^2g(E_F))$,
where $g(E_F)$ is the number of states per unit volume per unit
energy, estimated by Middendorff {\it et
  al.}\cite{middendorff_thermoelectric_1973}. In the present case, the
large $\kappa$ and small $m^*$ tend to make $a_B$ very large. For the
lowest carrier density samples investigated here $a_B\sim3$nm and
$\lambda_{TF}\sim 4$nm, which places is this material on the
metal-insulator boundary. The carrier density can also be reduced by
introducing foreign dopants which `drain' the excess carriers and pin
the chemical potential $\mu$ in the gap. For hydrogenic like
impurities this can be very effective, but in the present materials
impurity bands often form instead. At high enough impurity densities
the carriers may become Anderson localized. Such samples are
characterized by a high carrier density with very low mobility,
leading to a negative gradient in the temperature dependence of the
resistivity. This may be the case for example in
Bi$_{x}$Sb$_{y}$Pb$_{z}$Se$_3$ which has $\rho\sim 30$m$\Omega$cm yet
a carrier density $n_e\sim$5$\times
10^{18}$cm$^{-3}$\cite{kasparova_n-type_2005}. An Anderson insulator
is generally bad news for topological insulators, because even though
at zero temperature the bulk conductivity $\sigma=0$, at finite
temperature the transport may remain dominated by bulk hopping
mechanisms.

In conclusion, the present study reveals substantial agreement between
transport and ARPES measurements of the Fermiology of \bise, in
particular for samples with large carrier densities. However, for
samples with carrier densities approaching 10$^{17}$cm$^{-3}$,
discrepancies emerge as to the exact position of the Fermi level. We
have confirmed the bulk nature of the transport by the thickness
dependence of the Hall effect, resistivity and mobility. Furthermore
SdH data is highly consistent between different samples from the same
batch. Interestingly, the carrier densities measured here are an order
of magnitude smaller than those of the topological insulators recently
reported in the literature\cite{hsieh_observation_2009,chen_experimental_2009,hsieh_tunable_2009,
  peng_aharonov-bohm_2009, checkelsky_giant_2009}. ARPES
and STM have been invaluable tools in revealing the physics of
topological insulators, providing compelling evidence for the presence
of the topologically protected Dirac surface state. The present
results should stimulate further theoretical work as to the
consequences of the coexistence of bulk and surface states in a single
sample as well as innovation in novel ways to fabricate these
materials so the bulk state can be cleanly eliminated.

We would like to thank D. Goldhaber-Gordon, J. R. Williams, X. Qi,
S.-C Zhang, K. Lai, J. Koralek, J. Orenstein and T. Geballe for
useful discussions.  Work was supported by the U.S. DOE, Office of
Basic Energy Sciences under contract DE-AC02-76SF00515.

\bibliographystyle{apsrev}
\bibliography{bulkFSBi2Se3_arxiv}

\end{document}